\documentclass[11pt]{fizik}

\usepackage{hyperref}
\hypersetup{
colorlinks=true,
urlcolor=blue,
citecolor=blue}
\usepackage[all]{xy,xypic}
\usepackage{amsfonts,amssymb,amsmath,amsgen,amsopn,amsbsy,theorem,graphicx,epsfig}
\usepackage{eufrak,amscd,bezier,latexsym,mathrsfs,enumerate,multirow}
\usepackage[utf8]{inputenc}
\usepackage[english]{babel}
\usepackage[dvipsnames]{xcolor}
\usepackage{academicons}
\definecolor{orcidlogocol}{HTML}{A6CE39}
\usepackage[pagewise]{lineno}
\usepackage{epstopdf}
\usepackage{xspace}
\usepackage{color}
\usepackage{units}
\usepackage{slashed}
\usepackage{gensymb}
\usepackage{booktabs}
\usepackage{xcolor}
\usepackage{setspace}
\usepackage{physics}
\usepackage{nicefrac}
\usepackage{bbold}
\usepackage{accents}
\usepackage{mathtools}
\def\4u{4U~1608--52}

\usepackage[nolist,printonlyused]{acronym}
\begin{acronym}[TDMA]
 \acro{LMXB}{Low mass X-ray binary}
 \acrodefplural{LMXB}{Low mass X-ray binaries}
 \acro{AMXP}{accreting millisecond X-ray pulsar}
 \acrodefplural{AMXP}{accreting millisecond X-ray pulsars}
 \acro{PCA}{proportional counter array}
 \acro{ASM}{all sky monitor}
 \acro{RXTE}{the Rossi X-ray timing explorer}
 \acro{MAXI}{monitor of all sky X-ray image}
 \acro{Swift/XRT}{the Swift gamma-ray burst mission/X-ray telescope}
 \acro{Swift}{the Swift gamma-ray burst mission}
 \acro{XSPEC}{an X-ray spectral fitting package}
 \acro{PCU}{proportional counter unit}
 \acro{ISS}{international space station}
 \acro{MSP}{millisecond pulsar}
 \acro{NS}{neutron star}
 \acro{BH}{black hole}
 \acro{FFT}{Fast Fourier Transform}
 \acro{ML}{maximum likelihood}
\end{acronym}

\newcommand\aap{A\&A}                % Astronomy and Astrophysics
\newcommand\aapr{A\&ARv}             % Astronomy and Astrophysics Review (the)
\newcommand\apj{ApJ}                 % Astrophysical Journal
\newcommand\apjl{ApJ}                % Astrophysical Journal, Letters
\newcommand\apjs{ApJS}               % Astrophysical Journal, Supplement
\newcommand\mnras{MNRAS}             % Monthly Notices of the Royal Astronomical Society
\newcommand\na{New~Astron.}          % New Astronomy
\newcommand\nar{New~Astron.~Rev.}    % New Astronomy Review
\newcommand\pasj{PASJ}               % Publications of the Astronomical Society of Japan
\yil{}
\vol{}
\fpage{}
\lpage{}
\doi{}
\title{
Outburst Characteristics of Transient Low Mass X-ray Binary \4u}            %%% Insert your title here, do not remove } symbol
\author[GÜNGÖR]{                                                                                     %%% Insert abbreviated author names, to be seen at each 
\textbf{Can GÜNGÖR$^{1,2}$\thanks{gungor.can@istanbul.edu.tr}}\\ %%% Insert names of the authors explictly
$^{1}$\.{I}stanbul University, Science Faculty, Department of Astronomy and Space Sciences, Beyaz{\i}t, 34119, \\\.{I}stanbul, Turkey \\               %%% within { } symbols after \thanks, write down your correspondance e-mail adres (only one author is enough)
$^{2}$\.{I}stanbul University Observatory Research and Application Center, Beyaz{\i}t, 34119, \.{I}stanbul, Turkey\\                    %%% Insert affiliations of the authors, arrange numbering accordingly
\\ [1.8em]

\rec{.. 2025}
\acc{.. 2025}
\finv{.. 2025}
}

\newcommand{\bc}{\begin{center}}
\newcommand{\ec}{\end{center}}

\renewcommand{\phi}{\varphi}

\begin{document}

\maketitle

\begin{abstract} We present a deep study of the long-term X-ray light curve of \4u 
by investigating the fast rising exponential decay (FRED) outbursts, 
low intensity state (LIS) and quiescent intervals.
By calibrating the onset times of the outbursts, we identify three distinct classes for the FRED-type events:
\textit{(i)} the long-high outbursts, exceeding $\sim$50~d in duration with peak count rates above $\sim$40~cnt~s$^{-1}$;
\textit{(ii)} the short-medium outbursts, with durations of $\sim$20~d and peak count rates of $\sim$30--50~cnt~s$^{-1}$; and
\textit{(iii)} the short-low outbursts, also lasting $\sim$20~d but reaching only $\sim$20--30~cnt s$^{-1}$ at peak.
We, furthermore, examine the relation between pre-outburst duration and the peak $\&$ integrated count rates of the upcoming outburst.
We show that outbursts following longer quiescent periods tend to be more energetic.

\end{abstract}

\section{Introduction}
\label{intro}

Low mass X-ray binaries (LMXBs) are binary systems composed of a compact object, a black hole or a neutron star (NS),
and a low mass (M$_\ast \lesssim$ M$_\odot$) late type companion, also called the donor star since it is the source of the accumulating material.
The late type companion fills its Roche lobe, and material flows from its surface toward the Roche lobe of the compact object.
Due to the angular momentum of the transferred material and the system’s geometry, 
the material begins to orbit the compact object, eventually forming an accretion disk (see \cite{lewin95,bahramian23} for comprehensive reviews). 
Based on the temporal properties and the color-color diagram behavior, they are classified into two groups: Z and atoll sources.
Z sources are bright, mostly steady sources, while Atoll sources are typically fainter and may show transient nature
with unpredictable outbursts following long quiescent periods.

\4u is one of the most-studied transient LMXB composed of a NS and an optical counterpart (QX Nor).
It was first discovered as a transient source via the observation starting in the early 1970s \cite{belian76} 
and cataloged by UHURU satellite survey in 1978 \cite{forman78}.
The distance of the source is estimated as in the range of 2.9 $\lesssim$ d $\lesssim$ 4.5 kpc \cite{galloway08,poutanen14}, most likely 4.0~kpc \cite{ozel16}.
The orbital period of the system was estimated to be 0.5370 days via the systematic variability in the I-band optical observations \cite{watcher02}.
The spin frequency of the NS associated with the binary system is 619 Hz \cite{hartman03}.
\4u is very similar to another well studied LMXB system, Aql X-1, based on its long-term X-ray light curve 
with its very energetic and almost cyclic outbursts annually.
These sources are also very similar with their burst activity in X-ray \cite{campana98}.
The initial classification of outbursts using the X-ray light curve of Aql X-1 categorized energetic events into two types, FRED and LIS \cite{maitra08}.
FRED type outbursts having very steep rising stages and slow decays, are explained via the disk instability model \cite{chen97,lasota01}
while LISs are periods with higher energy levels than quiescent with a more chaotic X-ray light curve.
Another classification, in the literature, based on the rational intensity evolution in two energy bands, below and above 15.0 keV, 
splits outburst into two groups; Slow-type (S-type) and Fast-type (F-type) \cite{asai12} in which this study includes both Aql X-1 and \4u.
Using about seven years of data (1996–2003) obtained with the All Sky Monitor (ASM) aboard the Rossi X-ray Timing Explorer (RXTE), 
the long-term activity of \4u was analysed, revealing a wide variety of outburst morphologies with different durations and intensities, 
broadly consistent with the disk instability model \cite{simon04}. 
In this study, quasi-regular recurrence times were identified together with a trend of increasing peak intensity from one outburst to the next, 
although no clear correlation between peak intensity and outburst duration was found.
More recent monitoring studies have further confirmed that \4u shares with Aql X-1 the characteristic of exhibiting recurrent, 
nearly cyclic outbursts with comparable energetics.
An intriguing classification of Aql X-1 was reported by Güngör et al. (2014) \cite{gungor14} (hereafter G14), 
based on the pattern of the smoothed X-ray light curve during outbursts. In 
this study, three distinct classes were identified according to the peak count rates and durations of the outbursts: 
long-high, medium-low, and short-low.
Building on this framework, Güngör et al. (2017) \cite{gungor17a} (hereafter G17) examined the relation between 
the pre-outburst quiescent duration (i.e., the waiting time before an outburst) and the peak count 
rate, which serves as a measure of the released energy, suggesting that outbursts following longer waiting times tend to be more energetic.
Motivated by these findings, we extend this line of research to \4u, one of the most active and well-studied transient neutron-star LMXBs, 
in order to examine whether a similar classification scheme and quiescent duration/energetics relation can be established for this source.

In the recent catalogue of outbursts of NS conatining LMXBs \cite{heinke05}, \4u is identified as one of the best-monitored systems,
with dozens of outbursts detected across multiple X-ray monitoring missions.
The catalogue reports recurrence times of roughly annual to multi-year scales, large variations in outburst durations and peak intensities,
and a diversity of morphological types consistent with disk instability models.
Additional outputs include detailed tabulations of start and end times, peak fluxes and recurrence statistics for each event.
These results highlight the necessity of a systematic outburst classification scheme in order 
to interpret the wide range of behaviours observed in \4u and to enable consistent comparisons across different NS-LMXBs such as Aql X-1.
The outbursts of the source are individually investigated via spectral modelling in the literature \cite{gierliski02,armas17},
which show that the spectrum can be successfully modelled via a model including a disk blackbody from the accretion disk, 
a blackbody component from the NS surface or boundary layer, and a thermal Comptonization continuum.
Spectral analysis shows that the broadband spectra during the 2016 and 2020 outbursts are well 
described by a disk blackbody plus thermal Comptonization model,
indicating that the inner disk radius remained nearly constant ($\sim$22-27 km) 
while the accretion rate was higher during the 2016 outburst compared to 2020 \cite{bhattacherjee24}.
It has been shown that LMXBs, particularly compact systems with hydrogen-poor and helium or carbon/oxygen rich accretion disks, 
develop thermal–viscous instabilities under disk
instability model combined with irradiation, giving rise to recurring outbursts \cite{hameury16}.
Therefore, understanding the spectral evolution of different outbursts is crucial for 
constraining the accretion geometry and the interplay between the disk, boundary layer, and corona.

This paper presents the results of our comprehensive investigation of the long-term X-ray light curve of \4u.
The methodology adopted in this study is described in \autoref{method}.
Specifically, in \autoref{class}, the details of the data reduction and comparison/classification procedures for FRED type outbursts are given,
while the relationship between the pre-outburst quiescent duration and the peak count rate is presented in \autoref{duration}.
Our findings are discussed in the context of previous studies in \autoref{discuss}.
Finally, our conclusions are presented in \autoref{conclusion}.

\begin{figure}[ht!]
\centering
\includegraphics[width=15cm]{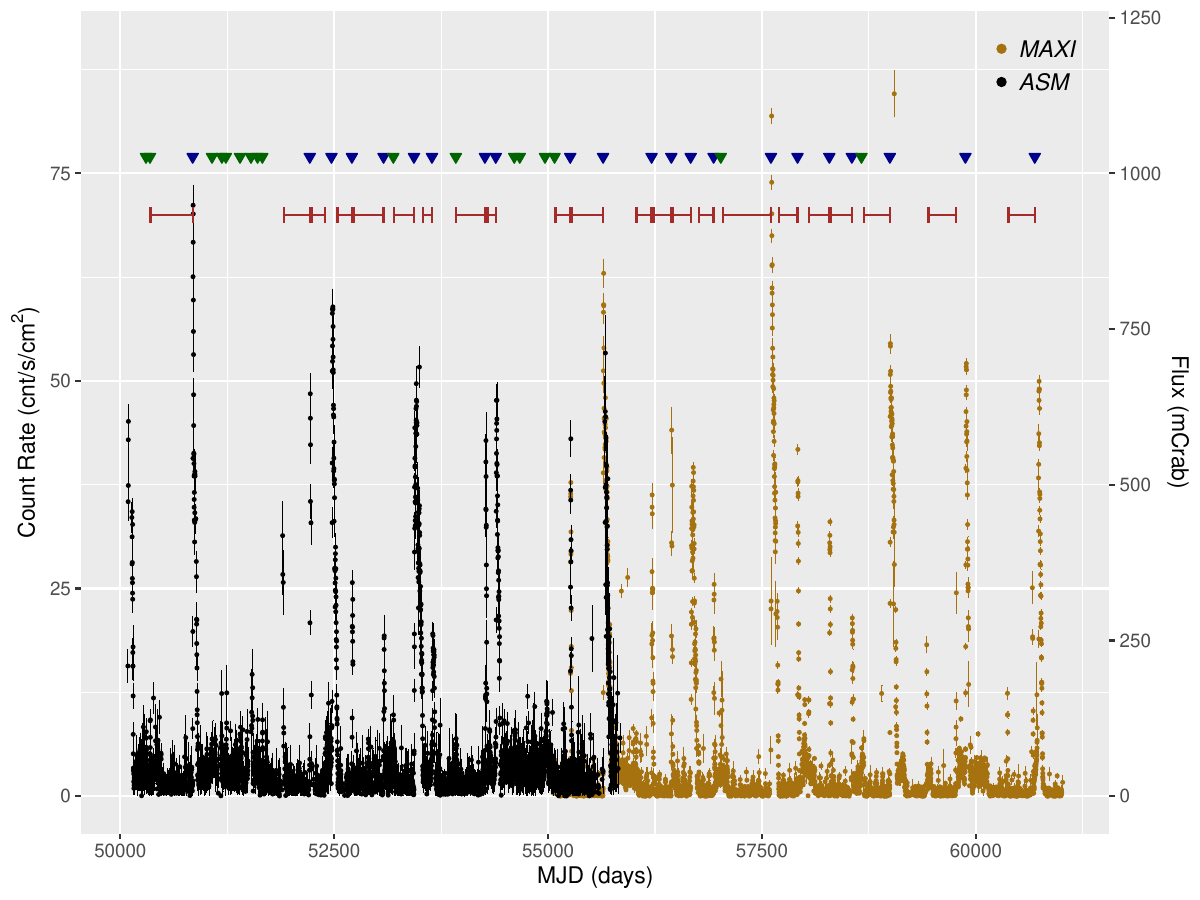}
\caption{The long-term light curve of \4u}
\label{figure1}
\end{figure}

\section{Methodology and Results}
\label{method} % label of section 2
\4u is a transient LMXB with its energetic outbursts and complicated X-ray light curve composed of FRED and LIS type events.
The long-term light curve of \4u is given in \autoref{figure1} created by using the whole daily average data of monitor of all sky X-ray image (MAXI) aboard international space station 
in 2.0-20.0 keV energy band and the sum band \cite{maxi}, 1.5-12.0 keV, daily average
count rates of the data obtained from RXTE/ASM \cite{asm}.
The FRED type outbursts and the LIS type events investigated in this study are highlighted with upside-down dark blue triangles and dark green triangles, respectively,
while brown horizontal lines indicate the pre-outburst durations.
The calibration constant between the two instruments was obtained to be 21.0 using the data from the outbursts in March 2010 and March 2011,
which were observed by both missions. 
This value is consistent with the one presented in G14, which was derived from the data of two outbursts of Aql X-1.
Therefore, the count rate values of MAXI were multiplied by this calibration constant to make the measurements from the two detectors directly comparable.
The upside-down dark blue and dark green triangles indicate the FRED and LIS type outbursts included in the sample set of this study, 
while the horizontal lines preceding the outbursts represent the pre-outburst quiescent periods.

\subsection{Classification of Outbursts}
\label{class}

We conduct a detailed examination of the entire long-term X-ray light curve.
As can be seen from \autoref{figure1}, the light curve of \4u exhibits considerable complexity,
arising from the diversity in the morphology of the FRED type outbursts as well as the presence of LIS events and
their combinations with FRED outbursts, reflecting the complex accretion behavior of the system.
Due to this complexity, performing a purely blind search is challenging and may result in overlooking real variations.

We first determined the quiescent level of the light curve and the
corresponding standard deviation during quiescence for the ASM and
MAXI data independently. We adopted a threshold defined as 5$\sigma$
above the mean of the quiescent state. For the ASM data, this
corresponds to 4.235~cnt~s$^{-1}$, based on a mean of 1.205~cnt~s$^{-1}$
and a standard deviation of 0.606~cnt~s$^{-1}$, while for the MAXI
data the threshold is 0.139, based on a mean of 0.019~cnt~s$^{-1}$ and a standard
deviation of 0.024~cnt~s$^{-1}$.
To eliminate false triggers due to statistical fluctuations, the
beginning of an event is identified when five consecutive data points
exceed the threshold. Similarly, the termination of an outburst is
established when five consecutive points fall below the threshold.
This approach provides a consistent and robust method for the automated
identification of the onset and end of both FRED and LIS type events.

\begin{figure}[ht!]
\centering
\includegraphics[width=15cm]{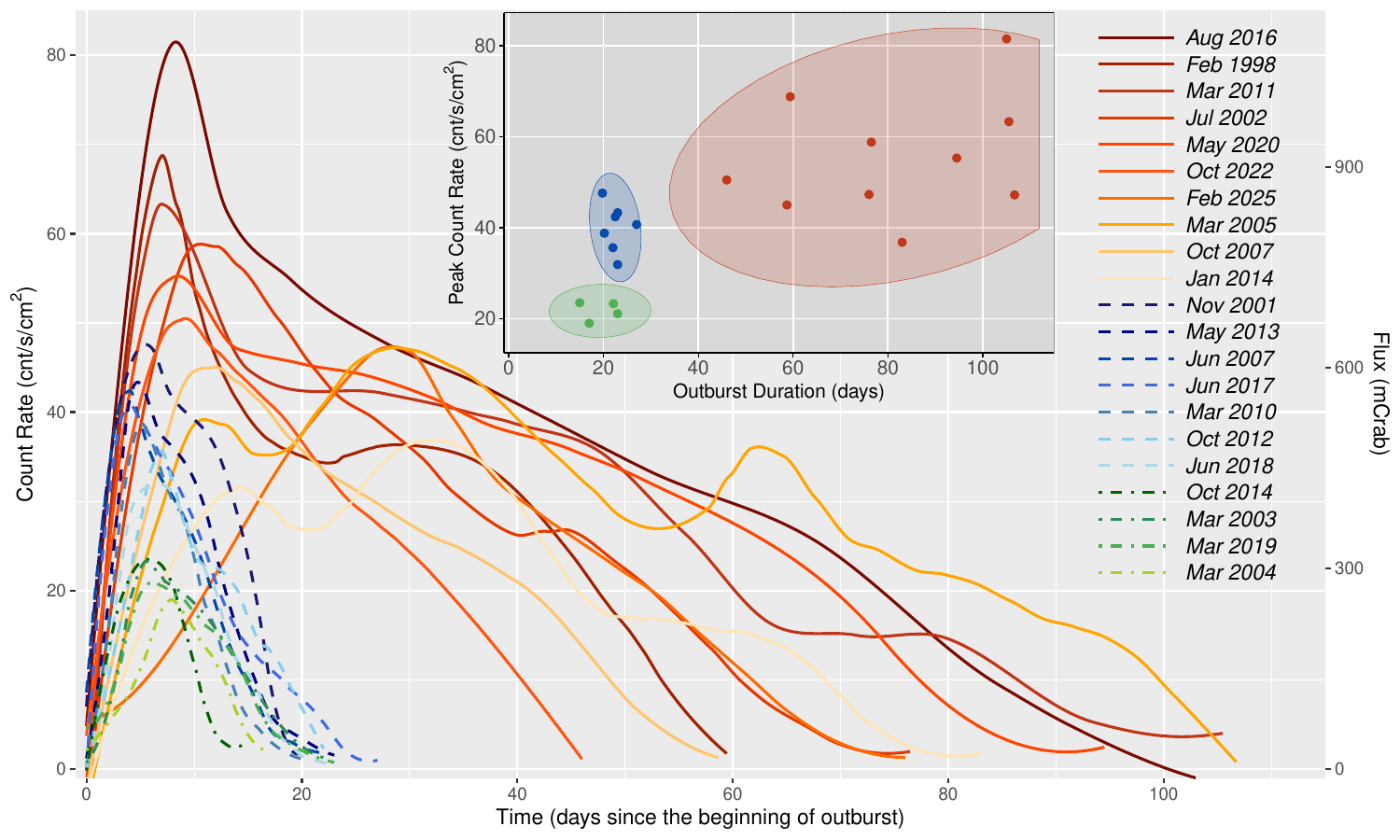}
\caption{The smoothed light curves of the FRED type outbursts of \4u calibrated to their onsets. 
The inner panel shows the relation between the pre-outburst duration and the peak count rate of each outburst, 
illustrating the separation of the long-high, short-medium, and short-low classes as revealed by the GMM fits.}
\label{figure2}
\end{figure}

All outburst onsets were calibrated by setting their starting time to zero.
Following a similar procedure to that in G14, we smoothed the light curves of FRED type outbursts using the s’Bézier method \cite{bezier}
to achieve a clearer visualization of the patterns and calibrated them to their onsets to enable a more comprehensive analysis.
Following the construction of the smoothed outburst profiles shown in \autoref{figure2} (main panel), we tested whether the apparent diversity in outburst duration and peak count rate reflects the presence of 
discrete groups rather than a continuous distribution. To this end, we modeled the distribution in the outburst duration vs peak count rate space by using Gaussian mixture models (GMMs) 
and applied a likelihood ratio test to compare 2 and 3 component models, corresponding to 2 class and 3 class approximations of the outburst population.
The test yields marginal statistical support for a three-component model (p $\approx$ 0.08), 
indicating that a 3 class description provides an improved representation of the data given the limited sample set. The resulting classification is illustrated in \autoref{figure2} (inner panel) using elliptical 
regions corresponding to the 85$\%$ confidence level.
We therefore identified three different outburst classes based on the peak count rate and the outburst duration:\\
\textit{(i)} the \textit{long-high} outbursts extending beyond $\sim$50~d with peak count rates above $\sim$40~cnt~s$^{-1}$;\\
\textit{(ii)} the \textit{short-medium} outbursts lasting $\sim$20~d with peak count rates of $\sim$30-50~cnt~s$^{-1}$; and\\
\textit{(iii)} the \textit{short-low} outbursts with a similar $\sim$20~d duration
but slightly lower peak count rates in the range of $\sim$20–30~cnt~s$^{-1}$.\\
The three outburst types --the long-high, short-medium, and short-low classes-- are represented
by solid, dashed, and dash-dotted lines with shades of red, blue, and green, respectively.
The downward order of the labels in \autoref{figure2} indicates the sequence of peak 
count rates within each outburst class.

\subsection{Pre-outburst Duration vs Peak $\&$ Integrated Emission}
\label{duration}
The search methodology based on the statistics from the daily averaged MAXI and ASM data also allows the identification of outburst onsets and terminations.
Accordingly, the pre-outburst quiescent periods are determined by the same automated search algorithm.
The smoothed X-ray light curves provide the peak count rates, which have commonly been used as indicators of the outburst strength.
It should be noted, however, that the time integral of the smoothed light curve provides a more representative measure of the total emitted energy during an outburst.
While we initially followed the same approach as in G17 to ensure direct comparability with previous studies of Aql~X-1,
we have now extended our analysis by computing the integrated count rate over each outburst.

\begin{figure}[ht!]
\centering
\includegraphics[width=15cm]{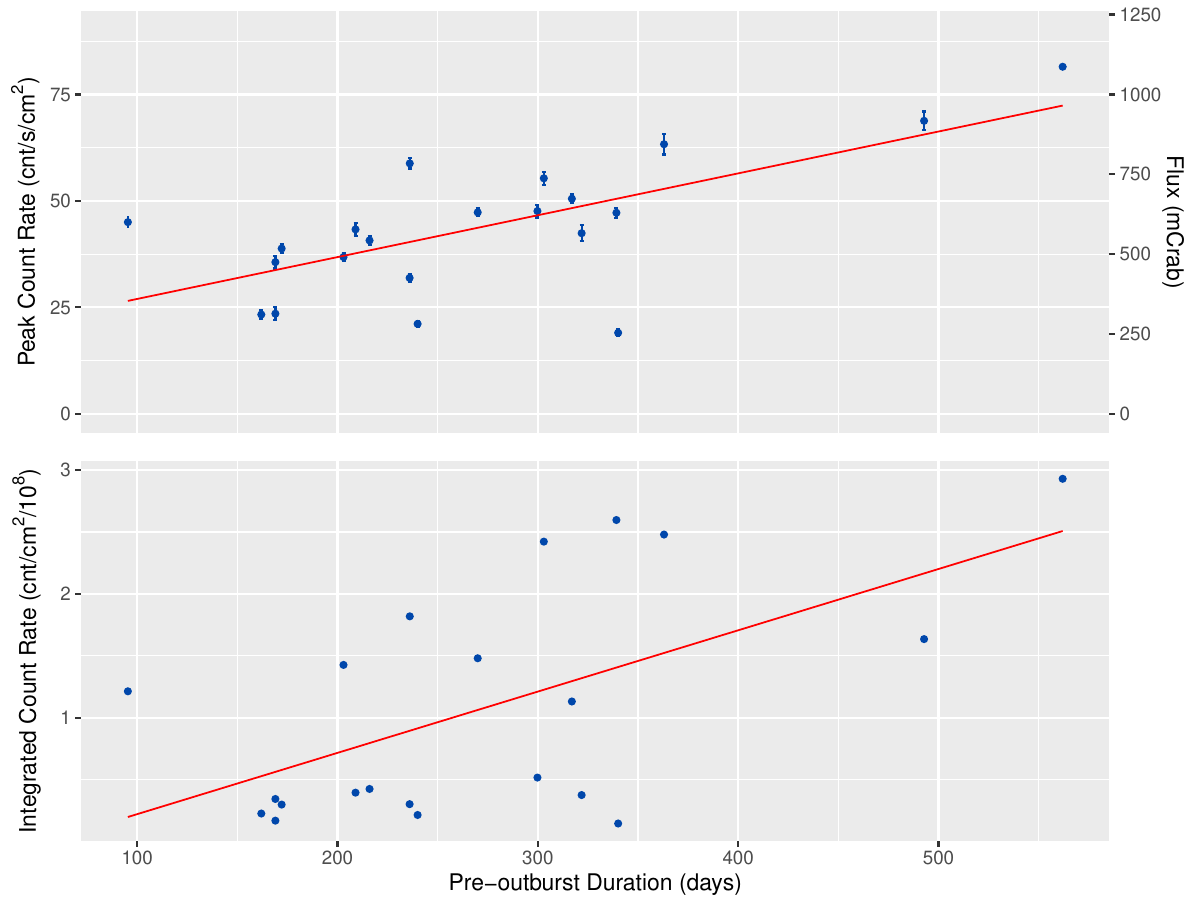}
\caption{Relation between the pre-outburst duration and the peak count rate (upper panel) and the integrated count rate over the outburst (lower panel) for FRED-type events of \4u.
The red line represents the best-fit linear trend to the data.}
\label{figure3}
\end{figure}

The relation between the pre-outburst duration and the peak count rate is shown in \autoref{figure3} (upper panel) for FRED-type outbursts,
together with the best-fit linear trend indicated by the red line.
To quantify this relation, we applied both Pearson and Spearman correlation analyses, as well as a linear regression fit.
We find a statistically significant positive correlation between the pre-outburst duration and the peak count rate
(Pearson $r = 0.68$, $p = 6.6\times10^{-4}$; Spearman $\rho = 0.56$, $p = 8.1\times10^{-3}$),
with a best-fit slope of $0.098 \pm 0.024$~cnt~s$^{-1}$~cm$^{-2}$~day$^{-1}$,
a scatter of $\sigma = 12.0$~cnt~s$^{-1}$~cm$^{-2}$, and $R^{2}=0.47$.

The corresponding relation between the pre-outburst duration and the integrated count rate is presented in the lower panel of \autoref{figure3}.
Applying the same statistical analysis, we again find a significant positive correlation
(Pearson $r = 0.59$, $p = 4.8\times10^{-3}$; Spearman $\rho = 0.50$, $p = 2.2\times10^{-2}$).
The linear fit yields a slope of $(4.9 \pm 1.5)\times10^{-3}$~(cnt~cm$^{-2}$)/$10^{8}$~day$^{-1}$,
with a scatter of $\sigma = 0.77$~(cnt~cm$^{-2}$)/$10^{8}$ and $R^{2}=0.35$.

We examined a possible correlation based on the idea that the accumulated mass in the accretion disk is proportional to the time
required to refill the reservoir.
In the case of Aql~X-1, G17 showed that there is no clear correlation between the pre-outburst duration and the peak count rate for LIS events,
as expected due to their variable light curve morphologies; LIS-type events are therefore not included in our analysis of \4u.

\section{Discussion}
\label{discuss}
This study demonstrates, for the first time, that although the X-ray light curve of \4u is highly complex, 
the FRED type outbursts exhibit three distinct classes:
the \textit{long-high} outbursts, lasting at least $\sim$50~d and reaching peak count rates above $\sim$40~cnt~s$^{-1}$;
the \textit{short-medium} outbursts, with durations of $\sim$30–40~d and peak count rates of $\sim$30–50~cnt~s$^{-1}$; and
the \textit{short-low} outbursts, lasting shorter than $\sim$30~d and reaching peak count rates of $\sim$20–30~cnt~s$^{-1}$.
The likelihood of an outburst being more energetic increases when it follows a prolonged quiescent period.
The dispersion observed in the pre-outburst duration versus peak count rate plane indicates that 
the waiting period is not the only parameter determining the properties of the subsequent outburst.
For a more precise investigation, the total energy released during an event could be quantified 
by integrating both the FRED type outbursts and LIS events.
While the current sample is sufficient to suggest a potential correlation, each newly observed outburst will further test and strengthen this hypothesis.
The reflection spectroscopy study of van den Eijnden et al. (2020) \cite{eijnden20} implicitly samples
all three outburst classes identified in this work, namely the Aug~2016,
Jun~2018, and Oct~2014 outbursts, which belong to the long-high,
short-medium, and short-low classes, respectively.
They showed that reflection features, such as the broad Fe~K$\alpha$ line
and the Compton hump, are prominent during brighter phases of outbursts,
but weaken or disappear as the source decays to lower luminosities.
This behaviour was interpreted as reflecting changes in the inner accretion
flow properties rather than being driven solely by observational effects.
Although their focus was on reflection spectroscopy,
their results provide an independent physical context supporting the view
that different outburst classes are associated with distinct accretion states
and geometries.

Aql X-1 provides a useful comparison for interpreting these results, as another transient NS-LMXB that exhibits a similar,
though not identical, classification of FRED type outbursts \cite{gungor14}.
In that study, three outburst types were introduced based on onset calibration: 
the \textit{long-high} outbursts extending beyond $\sim$50~d with peak count rates above $\sim$50~cnt~s$^{-1}$; 
the \textit{medium-low} outbursts lasting $\sim$20~d with peak count rates of $\sim$30–50~cnt~s$^{-1}$; and 
the \textit{short-low} outbursts with similar $\sim$20~d durations but slightly lower peak count rates of $\sim$20–30~cnt~s$^{-1}$.
The outburst trends exhibit different peak count rates and durations for Aql X-1 and \4u,
which is not surprising given their distinct physical and geometrical properties, such as orbital periods and disk structures.
For Aql X-1, G17 presented a possible relation between the pre-outburst waiting time and the peak count rate of the following outburst,
in contrast to the weak correlation previously reported in \cite{campana13}.
The study presented in this paper is a follow-up investigation on \4u, building upon G14 and G17,
which focused on another very similar transient NS-LMXB, Aql X-1.
In G14, 12 outbursts were used for the morphological classification, while in G17,
the relation between pre-outburst duration and peak count rate was examined using 14 FRED type outbursts.
In the present study, we applied the same methodology to \4u with an even larger sample consisting of 20 FRED type outbursts.
Our results strengthen the idea that outbursts are not indiscriminate, but rather follow a trend governed by certain,
yet unknown, discrete initial conditions that give rise to these three distinct classes.
As is well known, outbursts are explained by the disk instability model \cite{chen97, lasota01} and are directly related to the accretion rate.

To provide a physical scale for interpreting these outburst classes, we estimated 
the peak luminosities corresponding to the brightest events of each class.
Using the peak luminosity of the source during the 2018 outburst \cite{eijnden20},
we scaled the ASM-based peak count rates of the Aug~2016, Oct~2001, and Oct~2014 outbursts,
the most luminous examples of the \textit{long-high}, \textit{short-medium}, and 
\textit{short-low} classes, respectively.
Their estimated peak luminosities are 
$2.5\times10^{37}\,{\rm erg\,s^{-1}}$ ($0.12\,L_{\rm Edd}$),
$1.5\times10^{37}\,{\rm erg\,s^{-1}}$ ($0.07\,L_{\rm Edd}$), and
$7.4\times10^{36}\,{\rm erg\,s^{-1}}$ ($0.04\,L_{\rm Edd}$), respectively,
illustrating an approximate constraint for each class.
The duration of an outburst is fundamentally linked to how quickly matter is
redistributed through the accretion disk, making the viscous time scale a key
physical driver of the observed outburst time scales.
In irradiated disks, the effective cooling front is pushed outward, allowing the
disk to remain in a hot, viscous state for an extended period, thereby
lengthening both the rise and decay phases of an outburst \cite{king98}.
Modelling of Aql~X-1 has demonstrated that the decay profile, including the knee
feature, arises naturally from viscous evolution in an irradiated disk
\cite{lipunova22}, reinforcing the view that outburst durations and decay shapes
are primarily regulated by disk viscosity.
More recently, it has been shown that the typical outburst profiles of Aql~X-1—
including their characteristic durations and morphological evolution—can be
naturally reproduced through the coupled effects of disk irradiation, mass
accretion rate evolution, and viscous redistribution of matter within the
accretion disk \cite{coban24}.
These results indicate that the observed outburst durations in transient
NS-LMXBs are closely tied to the viscous time scale of the disk and to the
amount of mass accumulated during quiescence, providing a natural framework for
interpreting the correlation between waiting time and outburst energetics
observed in \4u.

The classification of outbursts presented in G14 and in this study suggests that the mass reservoir,
corresponding to the inner regions of the accretion disk, may exist in three distinct amount ranges.
G14 speculated that the inner disk radius may penetrate inward only during long-high 
outbursts, whereas during the short-medium and short-low classes it might remain close 
to the co-rotation radius ($R_{\rm co}$), potentially allowing a propeller-like regime with 
partial accretion, although this scenario is not supported by observational evidence in 
the current literature.
An alternative explanation could be related to the geometry of the accretion disk, in which different regions of the disk may be irradiated 
under varying morphological conditions.
It is natural to expect that, with a longer waiting time, the reservoir accumulates more material, leading to a more energetic subsequent outburst.
A very weak correlation between the peak count rate and the waiting time before an outburst was found in Aql X-1 \cite{campana13}.
G17 re-examined this hypothesis with a larger sample of the same source using smoothed X-ray light curves in a more systematic manner,
and found that outbursts following longer waiting periods tend to be more energetic, often reaching higher peak count rates.
Our results reinforce this hypothesis by showing the same correlation in another transient NS-LMXB, \4u.
It should be noted, however, that the statistically significant scatter of the data points around the linear trends in \autoref{figure3}, 
quantified by the moderate correlation coefficients and dispersion reported in \autoref{duration}, indicates that 
the pre-outburst quiescent duration is not the sole factor determining the peak intensity of a FRED-type outburst. 
This effect is likely to be particularly important in systems with complex X-ray light curves that include both LIS and FRED events, or their combinations.
Future studies involving a larger number of outbursts and applications to various transient sources will help 
to clarify the physical mechanisms underlying the different outburst classes.

\section{Conclusion}
\label{conclusion}
This study presents a systematic classification of FRED type outbursts in \4u, identifying three distinct morphological categories
and indicating that more energetic events tend to be preceded by longer quiescent periods.
The variability observed in the relationship between pre-outburst duration and peak and the integrated count rate suggests that
additional physical factors, such as accretion disk geometry, may influence the outburst energetics.
Furthermore, this work demonstrates that the classification framework previously proposed for Aql X-1 is also applicable 
to another transient LMXB, \4u, highlighting its physical relevance and broader applicability.
Continued monitoring and the identification of additional outbursts from these sources,
as well as similar analyses of other LMXBs, will further strengthen and refine the proposed classification.

\section*{Acknowledgment}
We thank the anonymous referees for their constructive comments and suggestions.
This work is partially supported by the Scientific and Technological Research Council of Turkey 
(TUBITAK) Grant No. \hbox{120F094}.
This research has made use of the MAXI data provided by RIKEN, JAXA and the MAXI team and
of results provided by the ASM/RXTE teams at MIT and at the RXTE SOF and GOF at NASA’s GSFC.
\\

%\bibliographystyle{JHEP}
%\bibliography{journal_bib}

\providecommand{\href}[2]{#2}\begingroup\raggedright\endgroup

\end{document}